# SECURE KEY EXCHANGE AND ENCRYPTION MECHANISM FOR GROUP COMMUNICATION IN WIRELESS AD HOC NETWORKS

S. Sumathy[1] and B. Upendra Kumar[2]

[1]School of Computing Sciences, VIT University, Vellore-632 014, Tamilnadu, India
`ssumathy@vit.ac.in`
[2] School of Computing Sciences, VIT University, Vellore-632 014, Tamilnadu, India
`upendra_mcs2@yahoo.com`

*ABSTRACT*

*Secured communication in ad hoc wireless networks is primarily important, because the communication signals are openly available as they propagate through air and are more susceptible to attacks ranging from passive eavesdropping to active interfering. The lack of any central coordination and shared wireless medium makes them more vulnerable to attacks than wired networks. Nodes act both as hosts and routers and are interconnected by Multi- hop communication path for forwarding and receiving packets to/from other nodes. The objective of this paper is to propose a key exchange and encryption mechanism that aims to use the MAC address as an additional parameter as the message specific key[to encrypt]and forward data among the nodes. The nodes are organized in spanning tree fashion, as they avoid forming cycles and exchange of key occurs only with authenticated neighbors in ad hoc networks, where nodes join or leave the network dynamically.*

## 1. INTRODUCTION

Ad hoc wireless networks are defined as the category of wireless networks that utilize multi-hop radio relaying and are capable of operating without the support of any fixed infrastructure and nodes communicate directly between one another over wireless channels [5]. As the wireless channels are openly available and propagate through the air, security in ad hoc networks is of primary concern [4]. In an ad hoc wireless network, the routing and resource management are done in a distributed manner in which all nodes coordinate to enable communication among them as a group. This requires each node to be more intelligent so that it can function both as a network host for transmitting and receiving data and as a network router for routing packets from other nodes. As ad hoc networks significantly vary from each other in many aspects, an environment-specific and efficient key management system is needed. To protect nodes against eavesdropping, the nodes must have made a mutual agreement on a shared secret or exchanged public keys. For very rapidly changing ad hoc networks the exchange of encryption keys may have to be addressed on-demand, thus without assumption about a priori negotiated secrets. The main advantage of ad hoc network is its economically less demanding deployment. In this work, a key exchange and encryption mechanism is presented where each node shares secret key only with authenticated neighbors in the ad hoc network thus avoiding global re-keying operations [1]. The proposed technique improves privacy protection and also takes care of anonymous routing performances in terms of network capacity using one-hop anonymous links with its neighbors. This methodology uses the MAC address of the nodes as an additional parameter for encryption as the message specific key among the nodes within the group. The encrypted data is appended with the ID of the receiving nodes before forwarding it in the network. So, the authorized destination node initially checks the ID of the received packet. If it matches, it further decrypts based on the neighborhood key as well as the message specific key. This mechanism requires the nodes to be organized in a spanning tree fashion as spanning tree is constructed with minimum distance which can cover all the nodes without forming a cycle.

## 2. KEY EXCHANGE AND ENCRYPTION TECHNIQUES

Active attacks involve actions performed by adversaries, for instance the replication, modification and deletion of exchanged data. External attacks are typically active attacks that are targeted to cause congestion, propagate incorrect routing information, prevent services from working properly or shut down them completely. External attacks can typically be prevented by using standard security mechanisms such as firewalls and encryption techniques. To secure group communication, nodes share a single symmetric key for encrypting and decrypting messages in existing systems. In the traditional group key exchange mechanism, if a new node joins or leaves, then the group key must be globally updated and distributed among the nodes in the group. This is referred to as group re-keying [1,6]. The disadvantage of this approach is that group re-keying needs access to a common server every time and can be complex and time consuming task. Access to a common server to generate a key every time a node joins or leaves in ad hoc environment is a time consuming process. Moreover it leads to consumption of resources in the ad hoc network. Group Key management is a subcategory of cryptography. Cryptography concerns itself with securing information so that unauthorized individuals cannot access and understand the messages sent. Key Management is essential for proper and secure distribution, creation and revocation of the keys used to secure messages [3]. To overcome the disadvantage of group re-keying, key exchange occurs only between the authenticated neighbors. Whenever there is a change in set of authenticated neighbors, a node must compute a new key and send this new key to all its authenticated neighbors [1]. Since the key is exchanged only with neighborhood nodes, the time taken to exchange the key is reduced considerably as well as authentication is also increased. For very rapidly changing ad hoc networks the exchange of encryption keys may have to be addressed on-demand, thus without assumptions about a priori negotiated secrets. Keys are securely exchanged in a Key update message after encryption using the traditional RSA algorithm.

### 2.1 SPANNING TREE CONSTRUCTION

Spanning tree of wireless nodes is constructed using NS2 simulator. Wireless nodes are plotted on the network animator at some x-coordinate and y-coordinate. The distance between each and every node is computed as follows:

$$\text{Distance } (i, j) = \overline{\sqrt{(x_j - x_i)^2 + (y_j - y_i)^2}} \qquad (1)$$

Spanning tree is constructed with minimum distance which can cover all the nodes without forming a cycle as in figure 4. The minimum spanning tree of wireless nodes is constructed and the spanning tree path is indicated in the network animator. The purpose of constructing spanning tree is that it is simple, cheap and an efficient way to connect terminals. Spanning trees are very important in designing efficient routing algorithms. In the ad hoc environment, after a spanning tree is built to connect a group of mobile devices, a packet can be always flooded to all members along the tree structure without loop and duplicated transmission [1]. As spanning tree maintains security associations only with neighbors, the proposed security scheme makes use of this mechanism.

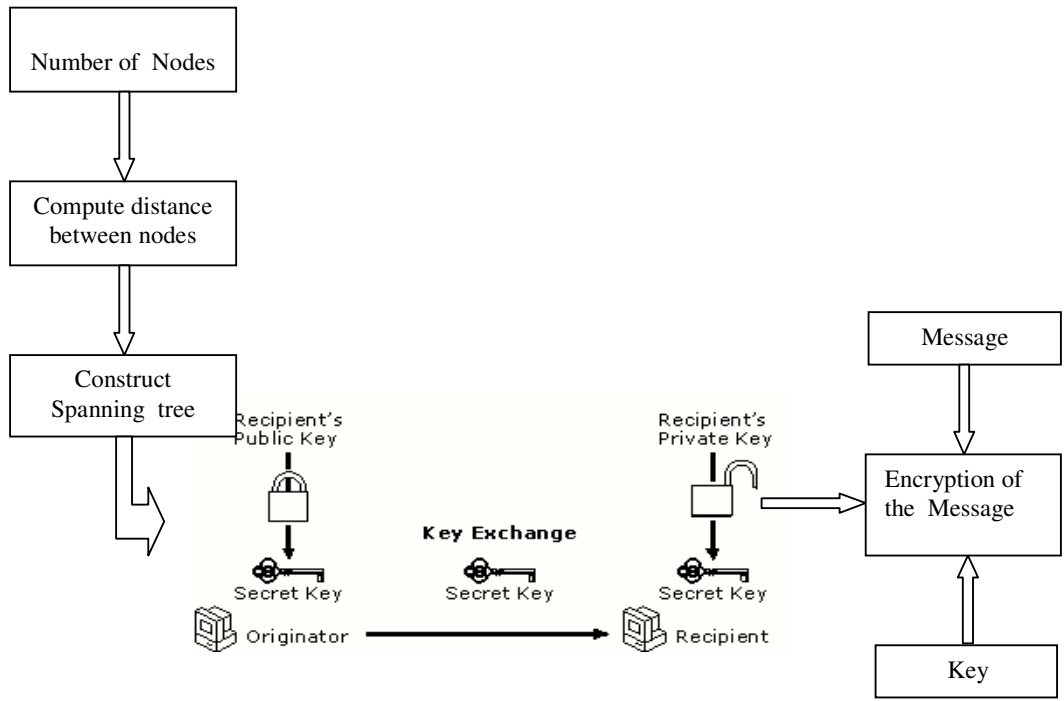

Figure 1. System Architecture [Encryption at the Source]

### 2.1.1 MINIMUM SPANNING TREE

Figure 2. shows the example scenario of the chosen undirected graph and using the spanning tree construction procedure the minimum spanning tree is constructed as in Figure 3. Every node exchanges data only with its neighbors. Information about the neighbors is maintained in a neighbor table. A node also maintains an adjacency table which contains the list of all nodes with which it can exchange messages.

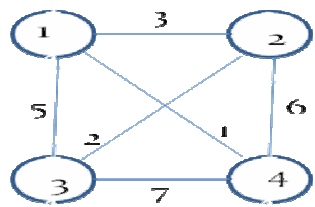

Figure 2. Undirected Graph

11

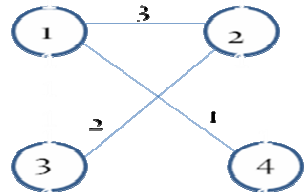

Figure 3.  Minimum Spanning Tree

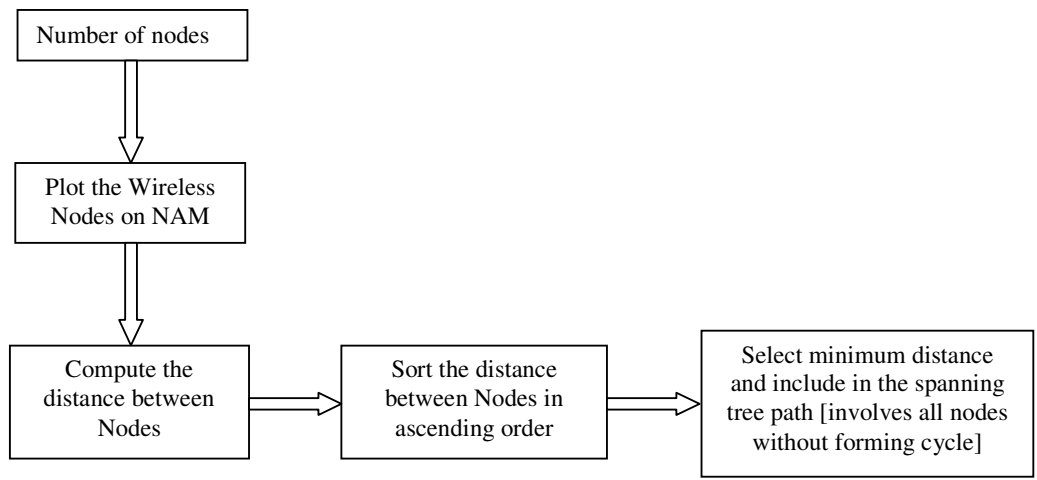

Figure 4.  Spanning Tree Construction

## 3. KEY EXCHANGE MECHANISM

The proposed security scheme consists of RSA key exchange mechanism and a novel encryption mechanism to provide security. Each node in the network has its own symmetric key called the Neighborhood key. To perform encryption and decryption each node must have access to other nodes neighborhood key. At source, Neighborhood Key is encrypted with the public key of the receiver and transmitted to the destination node. At destination, neighborhood key is decrypted with the node's own private key.

### 3.1 ENCRYPTION ALGORITHM

Each node has its own symmetric key called neighborhood key which is encrypted. Then, the message is encrypted using the message specific key which is the MAC address. Further, the message specific key is encrypted with neighborhood key. Then, the sender appends the destination nodes ID and transmits this message to its authenticated neighbors. Source Node A creates a Message Specific Key[MKey(M)]. Message is encrypted with Message Specific Key [$E_{MKey(M)}(M)$]. Further the Message Specific Key is encrypted with A's neighborhood key [$E_{NKey(A)}(MKey(M))$]. Then, the Destination node's ID is appended to the Ciphertext
[( $E_{NKey(A)}(MKey(M)$  $E_{MKey(M)}(M)$ ) Node ID(B)].

### 3.2 DECRYPTION ALGORITHM

At receiving end, if the ID of the node matches, then it is the intended recipient and decryption is performed with neighborhood key of sending node and the plain text message is obtained. As a next step, further decryption is done with the message specific key and the original message is obtained. If the node is not the intended recipient, it again re-encrypts the message with the neighborhood key and transmits to its authenticated neighbor nodes. The procedure is repeated until destination node is found and the original message is decrypted at the destination node. Two different symmetric encryption algorithms are used to encrypt the message and the neighborhood key with the message-specific key. The advantage of implementing two different encryption procedures is to make it to improve the security of the message being forwarded in the ad hoc network which is susceptible to more vulnerable attacks.

| | | | |
|---|---|---|---|
| Encrypted Message at node B | $E_{NKey(A)}(MKey(M))$ | $E_{MKey(M)}(M)$ | B |
| If B is the intended recipient decrypt with A's neighborhood Key | $D_{NKey(A)}(E_{NKey(A)}(MKey(M)))$ | $E_{MKey(M)}(M)$ | |
| Decrypt message with the obtained message key | $MKey(M)$ | $D_{MKey(M)}(E_{MKey(M)}(M))$ | |
| If B is not the intended recipient, ID of node C is appended | $E_{NKey(A)}(MKey(M))$ | $E_{MKey(M)}(M)$ | C |
| Decrypt the message key with A's neighborhood key | $D_{NKey(A)}(E_{NKey(A)}(MKey(M)))$ | $E_{MKey(M)}(M)$ | C |
| Re-encrypt the message with B's neighborhood key [assume spanning tree path as A---B--- | $E_{NKey(B)}(MKey(M))$ | $E_{MKey(M)}(M)$ | C |

Figure 5. Decryption Procedure

### 3.3 NEIGHBORHOOD KEY EXCHANGE PROCEDURE

Key exchange with only neighborhood nodes aims at reducing crypto-functions processing overheads occurred in a pure reactive approach. This neighbor detection scheme is identity-free and is carried over through a handshake process between any pair of neighbors. Handshaking procedure is basically carried over for key exchanges between a given node and its new detected neighbors. After the handshake procedure, each pair of nodes shares a chain of secret keys. HELLO messages are periodically sent to the nodes in the group. To forward the information the RREQ and RREP messages are used by each intermediate node to establish the route between the source and the destination nodes in the network.

## 4. SIMULATION RESULTS

The spanning tree is constructed using NS2 simulator. NS simulator is based on two languages: an object oriented simulator, written in C++, and OTcl (an object oriented extension of Tcl) interpreter, used to execute user's command scripts. A simulation script generally begins by creating an instance of this class and calling various methods to create nodes, topologies, and configure other aspects of the simulation.

As the nodes are organized in spanning tree topology in this security scheme, the nodes exchange keys and data only with its authenticated neighbors. This avoids expensive global re-keying operations when the membership in the network changes or when the network is partitioned. Figure 6. is a simulation output of wireless nodes plotted on NAM in the form of a spanning tree and the packets transferred between the nodes involved in the spanning tree path is obtained. Figure 7. and Figure 8. represents the simulation output of the throughput of the packets sent and the packets received.

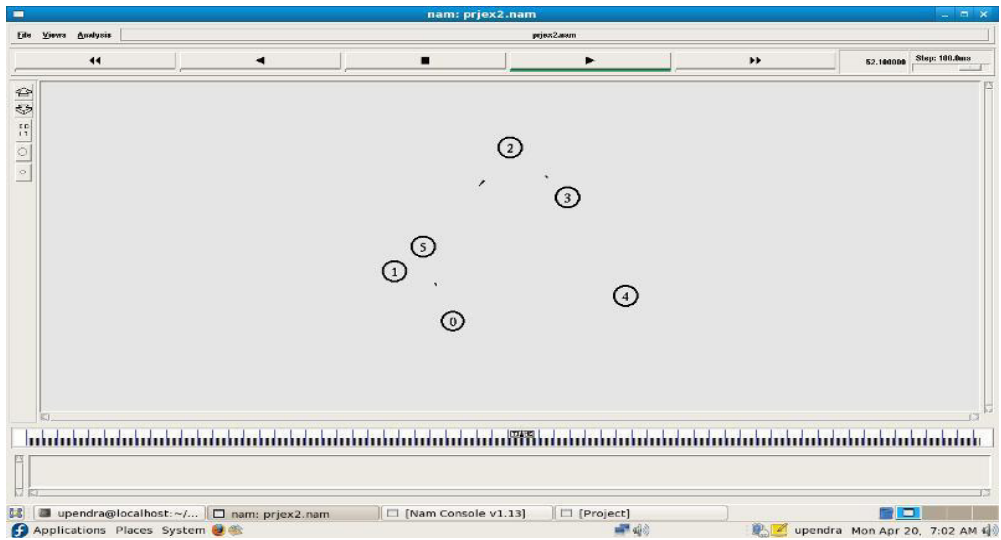

Figure 7. Packet transfer between nodes in spanning tree

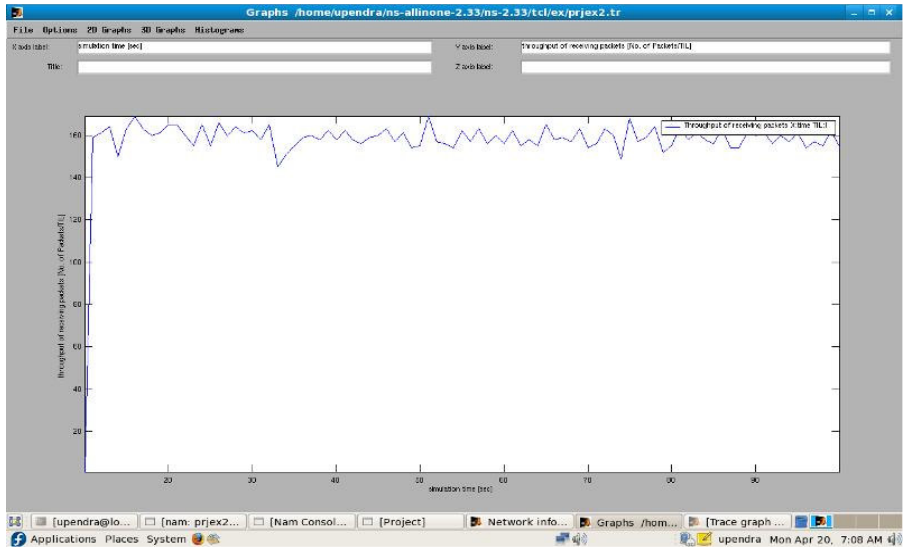

Figure 8. Throughput of Receiving Packets

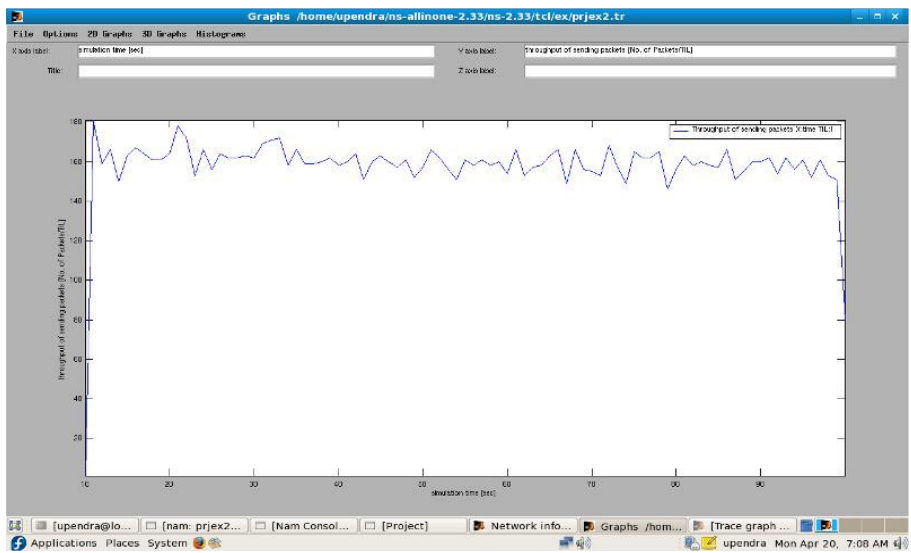

Figure 9. Throughput of Sending Packets

## 5. CONCLUSION AND FUTURE ENHANCEMENTS

The advantage of this security scheme is that since encryption is done twice with two different encryption schemes, one with neighborhood key and other with message specific key, more security is imposed. It ensures backward secrecy (a new member of network cannot access data transmitted before the member joined) and forward secrecy (a member cannot access the data that is transmitted after the member leaves the group)[1]. Whenever the topology changes with the inclusion or exclusion of a member, new neighborhood key is computed and is distributed to all authenticated neighbors.

A novel security scheme in ad hoc networks is presented which can address the security issues such as authentication, confidentiality and key management that would avoid global re-keying. The proposed scheme which aims at sender deniable encryption can be widely applicable for voting and auction protocols. This shall be applicable wherever group communications is to be established in a secured manner in an ad hoc environment. The future enhancement to the implementation of the proposed security scheme is to incorporate the key storage, message storage at the node level and compare the performance of spanning tree with and without the inclusion of features such as key storage and message storage.

## ACKNOWLEDGEMENTS

The authors would like to thank their Professer Dr. R.Saravanan, school of computing sciences, for his valuable suggestions in carrying out this work. Also wish to acknowledge the members of family and friends for their kind support.